# ATTENTION AND MISINFORMATION SHARING ON SOCIAL MEDIA

**[1,2] Zaid Amin, [*1]Nazlena Mohamad Ali, [3]Alan F Smeaton**

*[1,*]Institute of IR4.0 (IIR4.0), Universiti Kebangsaan Malaysia, Malaysia*

*[2] Faculty of Informatics Engineering,  Universitas Bina Darma, Palembang, Indonesia*

*[3] INSIGHT: Centre for Data Analytics, Dublin City University, Dublin 9, Ireland*

---

**Abstract:**

The behaviour of sharing information on social media should be fulfilled only when a user is exhibiting attentive behaviour. So that the useful information can be consumed constructively, and misinformation can be identified and ignored. Attentive behaviour is related to users' cognitive abilities in their processing of set information. The work described in this paper examines the issue of attentive factors that affect users' behaviour when they share misinformation on social media. The research aims to identify the significance of prevailing attention factors towards sharing misinformation on social media. We used a closed-ended questionnaire which consisted of a psychometric scale to measure attention behaviour with participants (n = 112). The regression equation results are obtained as: y=(19,533-0,390+e) from a set of regression analyses shows that attention factors have a significant negative correlation effect for users to share misinformation on social media.  Along with the findings of the analysis results, we propose that attentive factors are incorporated into a social media application's future design that could intervene in user attention and avoid potential harm caused by the spread of misinformation.

---

## 1. Introduction

Increasingly massive Internet growth has dramatically influenced the use of social media applications. According to Statista (2020), in July 2020, there were 4.57 billion active Internet users worldwide, and these active users cover 59% of the world's global population. The ability to share and openly discuss topics online has helped to give rise to social media applications' open development. The convenience that supports users in using social media applications further triggers them to share information more quickly and openly. However, the behaviour of sharing information on social media should be done only when the user is exhibiting attentive behaviour so that useful information can be consumed constructively, and misinformation can be identified and ignored.

When information is disseminated online without attentive behaviour, this can make misinformation spread and thus be misleading to recipients. As a result of the increasing spread of misinformation, this issue is becoming a focus of public attention. Various attempts have been made to overcome these challenges. One common approach widely used in designing online social media platforms that rely on algorithms to identifying fake stories as misinformation, as has been done by Facebook (Mosseri 2016). However, some research has justified that it is difficult to rely on efforts that rely on a machine and robotic approaches (Vosoughi et al., 2018). The actual "iceberg problem" that has not received much attention in research is the underlying psychological problem of user behaviour or a specific motive disseminating misinformation.

Sharing misinformation can trigger severe influences on people's own beliefs as well as causing reputational damage to the user who shared misinformation. The increase in shared misinformation on social media has now reached more people than the truth. Shared misinformation on social media negatively influences our responses to natural disasters, political events, and terrorist attacks (Vosoughi et al., 2018). Thus sharing misinformation can influence and threaten the stability of democratic life in a nation. Furthermore, managing the consequences of natural disasters and climate information, when the public is confused with misinformation, can cause the process of acquiring and sharing true knowledge and mitigating the effects of sharing misinformation, to be delayed (Cook et al., 2017).

Misinformation has affected the economic aspects of society, such as in-stock trading. An example of this is what happened on Twitter when Barack Obama was rumoured to have been injured in an explosion, which ended in a loss of $130 billion in stock value trading (Rapoza, 2017). Moreover, in 2014, a story about Ebola victims became viral, and even though the news was a lie, it was still shared online millions of times. These kinds of hoaxes happen every day in social media, and many of these motives only seek profits for the sake of attracting high traffic to a website (Hodson, 2015).

Regardless of the controversies around privacy, misinformation, and other negative features of online life, our world extends to comprise the Internet and social media. Roughly 45% of the world's population are social media users totalling nearly 3.8 billion people (Kemp, 2020). Increasingly we see that people are willing to share information online without reading it first, indicating their lack of attentive behaviour, when just reading a title of an article without understanding or verify its contents. In fact, the results of a study by Gabielkov et al. (2016) revealed that attentive user behaviour on social media is displayed when they share about 59% of Twitter links without even reading them.

This phenomenon also confirms that things such as the intensity of the news or information topic can influence users' behaviour to share information on social media (Bonchi et al., 2011). The quality of the information contents is not a mandatory requirement for online virality (Weng, 2012). Although the research conducted by Ghaisani et al. (2017) shows what motivates a person to share information on social media, not much is comprehended in particular about what factors influence a person in sharing information. We refer to (Bock, Zmud, Kim, & Lee, 2005), who show that information sharing behaviour is influenced by several factors such as beliefs, similar interests (social influence) and factors such as wanting to have an advantage in promoting themselves.

Studies that have examined information sharing between individuals have presumed that the behavioural routine of sharing information is habitually considered to be benefit-oriented. In the context of motivation, research conducted by Munar & Jacobsen (2014) describe that motivation in sharing information is divided into two parts, namely self-centered motivation and community-related motivation and this is something we explore further here.

This study aimed to determine the extent of the relationship between attention behaviour factors and misinformation sharing behaviour on social media. The significance of knowing the relationship between attentive behaviour and sharing misinformation will be a future reference point in the development of applications for information sharing, both in terms of the design and function of the application itself.

In this study, we aim to (1) determine the role of attention factors relating to user behaviour in sharing misinformation on social media and to describe the scope of how far the research topic has been developed; (2) define existing approaches which consolidate results, surveys, and methods through a statistical inferential outlook to assist businesses and organisations in designing appropriate social media applications.

By defining existing approaches based on data surveys, it is found that attention factors have a significant negative correlation effect for users when sharing misinformation on social media. We expect these findings will be useful as a basis for developing application designs on social media. Overall this paper is divided into six sections. Section 1 contains an introduction to the background of this research. Section 2 defines misinformation and discusses some of the related research. Section 3 discusses a survey to clarify the attention factors and data, obtained with cross-sectional studies from 112 respondents with a closed-ended questionnaire. Section 4 analyses survey results with a regression analysis approach, while the variables being analysed measure the relationships between variables x (attention) and y (misinformation sharing behaviour/MSB). Section 5 discusses the analysis results, and section 6 concludes the study and provides future direction in particular on this topic.

## 2.  Background Work

### 2.1. Attention

Attentive behaviour in a person is one of the fundamental phenomena that apply when a person interacts with their environment. Therefore, in the early stages of building an application for sharing social media information, it is essential for those who focus on designing Human-Computer Interaction (HCI) to examine attention and their interactive associations with action planning (Salihan, Nazlena, & Masnizah, 2017).

Attention is defined as a series of processes that allow a person to process information in a limited manner through all available information regarding cognitive abilities. Affirmed in different approaches, attention is a process of information processing inserted into working memory and provides a level of awareness. An extensive literature on selective attention, as described by (Jacko, 2012), states that there is a prominent role regarding selective attention when screening targets for information to process. This is also related to HCI when determining a stimulus's appearance, highlighting with colour and sound. The attention factor should not be considered a unitary function, but rather as a series of information processing activities related to perception, cognitive and motor skills that differ in each person and each situation.

According to Welsh et al. (2009) there are three important characteristics of attention: "(1) selective attention … only provide certain pieces of information to enter the limited processing system; (2) the focus of attention can be changed from one source of information to another; (3) attention can be separated within specific limits; one can selectively attend more than one source of information at a time".

According to Mcavinue & Habekost (2012) to estimate attentional selectivity and capability employing a standard based on the theory of visual attention, we must understand which allows the calculation of parameters that correlate with a selection of attention, perceptual threshold, short-term visual memory capacity, and processing capability and capacity.

The ideal duration of attention is still highly debated because it is very much task-dependent. How significant is the level of attention that is required, will vary depending on what the task demands are. How we practice our attention to various tasks depends on what the individual carries to the circumstances. Current developments in the subject of how selective attention interferes with perception and action and, in turn, how intentional acts influence attentional processes, are yet to make such a contribution to misinformation sharing behaviour.

*2.2. Attention and Information Sharing*

Weng et al. (2012) found that the attention factor is important in explaining how users behave on social media. The act of attentiveness plays a fundamental role, especially in reading tweets, browsing web pages, and responding to emails. Such online activities require mental effort and stress the human brain's capacity. As a result, in a given circumstance or task, the more (visual) stimuli that must be processed, the less likely the user will be to responding to stimuli because they have to share their limited attention with the incoming (visual) stimuli. This phenomenon is known as divided attention (Hodas and Lerman, 2012). Their study shows that limited attention and visibility are among the problems in disseminating information on social media networks. Divided attention also plays an important role in social contagion, the process of sharing information from person to person online through follower relationships.

Research on empirical studies discusses how cognitive users can respond to received stimuli such as a tweet (message) from a friend with limited visibility (the issue of screen size compatibility), which requires increased cognitive effort in responding to the tweet. In particular (Hodas and Lerman, 2012) found that users retweet information when the tweet is most visible in their visual interface, such as when the tweet is near the top of their Twitter feed.

Referring to Hodas and Lerman (2012), while the attentiveness factor is inherently a property of individual users and is limited for each task and situation, social media designers would influence users through their interface design choices. Interface design can manipulate user visibility to maximise user attention when consuming information on social media applications.

*2.3. Misinformation*

According to Wu et al. (2019), there are several terms related to misinformation that need to be described. These include spam (multiple email recipients), rumours, and fake news (in a modified news format), to the very similar term disinformation.

Misinformation and disinformation are false or inaccurate information. The main key difference between them lies in the maker's intent towards the sender and whether the information is intentionally misled. The term disinformation usually refers to cases that have an intentional motive, whereas misinformation is unintentional.

Throughout the discussion of their research, Wu et al. (2019) refer to misinformation as a general term to include all known false or inaccurate information (viral) and spreads on social media. Several concepts are discussed by Wu et al. (2019), such as misinformation, disinformation, spam, rumours, fake news. All of them believe that inaccurate messages can cause harm and various kinds of destructive effects through social media to the real environment, especially if there is no timely intervention such as dealing with terror, natural disasters, and health issues. In summary, Wu et al. (2019) regulate the various types of misinformation as follows:

1. Unintentionally Spread Misinformation:

In some cases, misinformation is unintentional to deceive the recipient. Users can easily contribute to sharing information because they easily trust the source of information. Information that is often received and easily disseminated is from close groups such as friends, family, colleagues, or users who influence social media. Instead of deceiving and taking advantage, they usually try to be the most up to date informing their social friends about a particular problem or situation.

2. Intentionally Spread Misinformation:

Some misinformation is deliberately spread to trick the recipient, which triggers intensive conversations and goes viral about misinformation and hoaxes. "There are usually writers and disseminators" organised behind a particular individual or business' popularity, who have clear objectives and agendas for gathering and promoting misinformation for profit.

Common examples of misinformation being deliberately spread include conspiracies, rumours, and fake news trending during the 2016 US Presidential Election. For example, a fake news writer, Paul Horner, has claimed credit for several fake news stories that went viral in 2017 (Wu et al. 2019).

## 3. Methodology

The research method conducted in this study was carried out systematically, starting from the determination of sampling, questionnaire design, analysis of results from the index of variables, and then proceeding with inferential statistical analysis. The purpose of defining this systematic method is that the resulting interpretation process is valid and reliable so that it can be recognised how significant is the factor of attention in sharing misinformation on social media. The attention variable is the independent variable (x), and the misinformation sharing behaviour variable is the dependent variable (y), as shown in Figure 1.

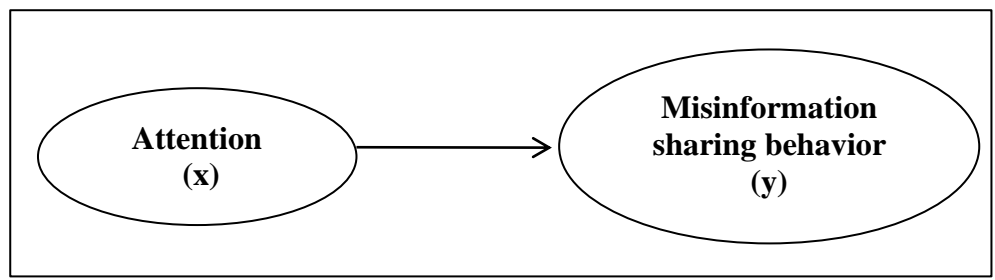

**Figure 1.** [Related variables]

Concerning the relationship between attention and misinformation sharing behaviour, we state the following hypotheses: H0: Attention does not influence misinformation sharing behaviour and H1: Attention influences misinformation sharing behaviour.

The data collection technique in this study used non-probability sampling with a purposive sampling method approach. The sampling method used pre-determined criteria such as age range and user activity on social media. The instrument applied is an online questionnaire. The poll provides questions to determine the users' answers about the relationship and the effect of attention behaviour in sharing misinformation on social media.

The questions consisted of demographic information, including gender, profession, age, education, and frequently used social media platforms. For every kind of information, the respondent is asked to fill his/her attention to sharing a specific kind of information on social media. Respondents are asked to choose one item that suited the most (5 Likert scales based). Analysis techniques that were used in the analysis include:

*a. Variable Index Analysis*

In this study, the type of descriptive analysis for the index of variables used is the analysis of index values, which is useful for obtaining a description of the respondents' perceptions of the questions raised. To calculate the index value, the following formula is used (Ferdinand, 2006). F1 is the frequency of respondents who answered 1; F2 is the frequency of respondents who answered 2, and so on until F5 for answer 5.

By using the three-box method criteria, as a basis for interpreting the index value is: Index Value = ((%F1x1)+(%F2x2)+(%F3x3)+(%F4x4)+(%F5x5))/5, with a range of criteria is 10,00–40,00 is low, 40,0 -70,00 is middle, and 70,01-100,00 is high.

*b. Inferential Statistical Analysis*

Refer to Garson (2012). The inferential statistical analysis includes instrument test (validity and reliability test), classic assumption tests include linearity test, normality test with Kolmogorov-Smirnov (K-S), heteroscedasticity test with the Glejser test, hypotheses (t-test) and R-squared. In general statistical analysis, OLS (Ordinary Least Squares) regression has the underlying assumptions rather than the research results. If the classical assumptions for linear regression are declared valid, then least squares can produce the best estimates.

## 4. Results

*4.1 Respondents' Demographics*

The respondents' data were collected between October 1, and November 30 2018 (8 weeks), and 112 answers from participants were collected. Demographic data are presented in Table 1, showing that the sample contains slightly more women (55.4%). The origin of the largest participating country is Indonesia at 92.9%. The most widely used social media platform is WhatsApp at 86.6%, and the age range of most participants is 30-39 years at 39.3%.

**Table 1**. [Respondent's demographics]

| Demographic data | | Number / Percentage | Demographic data | | Number / Percentage |
|---|---|---|---|---|---|
| **Gender** | Male | 44.6% | **Country of origin** | Indonesia | 92.9% |
| | Female | 55.4% | | Malaysia | 7.1% |
| **Age** | <17 years | 12.5% | **The most used social media** | Twitter | 11.6% |
| | 18 – 20 years | 2.7% | | WhatsApp | 86.6% |
| | 21 – 29 years | 33% | | Facebook | 50% |
| | 30 – 39 years | 39.3% | | Instagram | 55.4% |
| | 40 – 49 years | 8% | | YouTube | 24.1% |
| | 50 – 69 years | 4.5% | | | |
| | 60> years | 2.7% | | | |
| **Highest education** | Less than high school degree | 12.5% | | | |
| | High school degree | 15.2% | | | |
| | Diploma degree | 3.6% | | | |
| | Bachelor's degree | 40.2% | | | |
| | Master's degree | 28.6% | | | |
| | Postgraduate degree | 12.5% | | | |

*4.2 Variable Index Analysis Results*

Using the index variable analysis guidelines for predetermined variables, the results of index numbers for each variable of attention and misinformation sharing behaviour based on a questionnaire (q.x results) can be calculated as follows (see Table 2).

**Table 2**. [Variable Index Analysis Results]

| Attention (X) | Likert scale | | | | | Total % | Index % |
|---|---|---|---|---|---|---|---|
| | %F1 | %F2 | %F3 | %F4 | %F5 | | |
| q.1 | 17,9 | 9,8 | 20,5 | 30,4 | 21,4 | 327,6 | 65,52 |
| q.2 | 13,4 | 14,3 | 19,6 | 32,1 | 20,5 | 331,7 | 66,34 |
| q.3 | 9,8 | 4,5 | 6,3 | 12,5 | 67 | 422,7 | 84,54 |
| **Mean** | | | | | | | **72,1** |
| **MSB (Y)** | **Likert scale** | | | | | **Total %** | **Index %** |
| | %F1 | %F2 | %F3 | %F4 | %F5 | | |
| q.1 | 13,4 | 20,5 | 22,3 | 21,4 | 22,3 | 318,4 | 63,68 |
| q.2 | 18,8 | 23,2 | 26,6 | 20,5 | 8,9 | 271,5 | 54,3 |
| q.3 | 9,8 | 8,9 | 24,1 | 27,7 | 29,5 | 358,2 | 71,64 |
| q.4 | 10,7 | 11,6 | 26,8 | 20,5 | 30,4 | 348,3 | 69,66 |
| q.5 | 9,8 | 6,3 | 23,2 | 22,3 | 38,4 | 373,2 | 74,64 |
| **Mean** | | | | | | | **66,8** |

The study results obtained an average value of the attention variable index of 72%, and it can be concluded that many respondent responses regarding the attention variable are in the high category. With this average value, it can be seen that the attention of users greatly affects misinformation sharing behaviour on social media, and with the average value of the misinformation sharing behaviour variable index at 67%, it can be concluded that the majority of respondents' responses regarding their misinformation sharing behaviour on social media is in the high category.

*4.3 Instrument Test Results*

We conducted Pearson correlations between the various items. Validity testing was done by comparing the value of r arithmetic with the value of r table (n-2). The level of confidence was 95% (sig $\alpha$ = 0.05) and the degree of freedom (df) = n-2 so that the r table was 0.1548 (Garson, 2012). Validity test results for item x1.q1 ($\beta$ = 0.892, $p < 0.05$) is valid, item x1.q2 ($\beta$ = 0.864, $p < 0.05$) is valid, and item x1.q3 ($\beta$ = 0.669, $p < 0.05$) is valid. Validity test results for item y.q1 ($\beta$ = 0.749, $p < 0.05$) is valid, item y.q2 ($\beta$ = 0.602 $p < 0.05$) is valid, item y.q3 ($\beta$ = 0.774, $p < 0.05$) is valid, y.q4 ($\beta$ = 0.767, $p < 0.05$) is valid, and y.q5 ($\beta$ = 0.778, $p < 0.05$) is valid. Based on the results of validity testing carried out in Table 3, it was found that all items used to measure each variable were declared valid.

**Table 3**. [Validity test results for x & y items]

| | | x1.q1 | x1.q2 | x1.q3 | X | Results | | | y.q1 | y.q2 | y.q3 | y.q4 | y.q5 | Y | Results |
|---|---|---|---|---|---|---|---|---|---|---|---|---|---|---|---|
| x1.q1 | Pearson Correlation | 1 | .813** | .344** | .892** | Valid | y.q1 | Pearson Correlation | 1 | .389** | .434** | .468** | .433** | .749** | Valid |
| | Sig. (2-tailed) | | .000 | .000 | .000 | | | Sig. (2-tailed) | | .000 | .000 | .000 | .000 | .000 | |
| | N | 112 | 112 | 112 | 112 | | | N | 112 | 112 | 112 | 112 | 112 | 112 | |
| x1.q2 | Pearson Correlation | .813** | 1 | .283** | .864** | Valid | y.q2 | Pearson Correlation | .389** | 1 | .367** | .206* | .289** | .602** | Valid |
| | Sig. (2-tailed) | .000 | | .002 | .000 | | | Sig. (2-tailed) | .000 | | .000 | .030 | .002 | .000 | |
| | N | 112 | 112 | 112 | 112 | | | N | 112 | 112 | 112 | 112 | 112 | 112 | |
| x1.q3 | Pearson Correlation | .344** | .283** | 1 | .669** | Valid | y.q3 | Pearson Correlation | .434** | .367** | 1 | .524** | .525** | .774** | Valid |
| | Sig. (2-tailed) | .000 | .002 | | .000 | | | Sig. (2-tailed) | .000 | .000 | | .000 | .000 | .000 | |
| | N | 112 | 112 | 112 | 112 | | | N | 112 | 112 | 112 | 112 | 112 | 112 | |
| X | Pearson Correlation | .892** | .864** | .669** | 1 | | y.q4 | Pearson Correlation | .468** | .206* | .524** | 1 | .599** | .767** | Valid |
| | Sig. (2-tailed) | .000 | .000 | .000 | | | | Sig. (2-tailed) | .000 | .030 | .000 | | .000 | .000 | |
| | N | 112 | 112 | 112 | 112 | | | N | 112 | 112 | 112 | 112 | 112 | 112 | |
| | | | | | | | y.q5 | Pearson Correlation | .433** | .289** | .525** | .599** | 1 | .778** | Valid |
| | | | | | | | | Sig. (2-tailed) | .000 | .002 | .000 | .000 | | .000 | |
| | | | | | | | | N | 112 | 112 | 112 | 112 | 112 | 112 | |
| | | | | | | | Y | Pearson Correlation | .749** | .602** | .774** | .767** | .778** | 1 | |
| | | | | | | | | Sig. (2-tailed) | .000 | .000 | .000 | .000 | .000 | | |
| | | | | | | | | N | 112 | 112 | 112 | 112 | 112 | 112 | |

In this study, reliability testing was carried out by testing Cronbach Alpha (Garson, 2012), using the criteria if the Cronbach Alpha coefficient is greater than the significance level of 70% or p> 0.7, then the variable used is reliable. The reliability test results for item x ($\beta = 0.736$, $p > 0.7$) are reliable and for item y ($\beta = 0.787$, $p > 0.7$) is also reliable. Based on these findings, it can be seen in Table 4 that all variables have Cronbach Alpha values that meet the criteria of $p > 0.7$ so that it can be concluded that all variables in this study are reliable.

**Table 4.** [Reliability test]

| Reliability Statistics Results for X items | | |
|---|---|---|
| Cronbach's Alpha | N of Items | Results |
| .736 | 3 | **Reliable** |
| Reliability Statistics Results for Y items. | | |
| Cronbach's Alpha | N of Items | Results |
| .787 | 5 | **Reliable** |

*4.4 Classic Assumption Test Results*

The linearity test refers to the value of deviation from linearity. If $p > 0.05$ then there is a significant linear relationship between the independent and dependent variables. If the calculated F value < F table, then there is a significant linear relationship between the independent and dependent variables (Garson, 2012). Based on the significant value of the resulting analysis, the value of deviation from linearity Sig. is ($\beta = 0.102$, $p > 0.05$) and (F table= 1.628 < F value= 3.93), so from this result in table 5 it can be concluded that there is a significant linear relationship between the variable of attention (x) with misinformation sharing behaviour (y).

**Table 5.** [Linearity test results]

ANOVA Table

| | | | Sum of Squares | df | Mean Square | F | Sig. |
|---|---|---|---|---|---|---|---|
| sumY * sumX | Between Groups | (Combined) | 536.238 | 12 | 44.686 | 2.234 | .015 |
| | | Linearity | 178.002 | 1 | 178.002 | 8.899 | .004 |
| | | Deviation from Linearity | 358.236 | 11 | 32.567 | 1.628 | .102 |
| | Within Groups | | 1.980.253 | 99 | 20.003 | | |
| | Total | | 2.516.491 | 111 | | | |

The linear correlation depicted in Figure 2 shows that the scatter plot has a significant relationship with the negative correlation between the attention variable (x) and the misinformation sharing behaviour (y) variable.

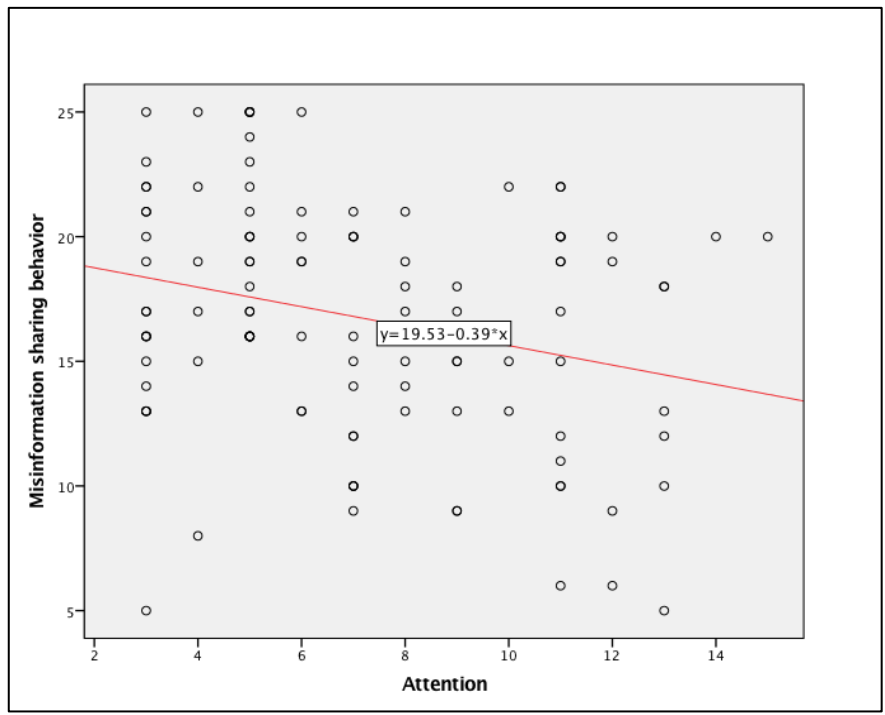

**Figure 2.** [Scatter plot shows a negative correlation between variables x and y]

Testing for normality in this study was conducted by examining data distribution, which is the normal probability distribution. The normality test results with the normal p-plot can be seen in Figure 3, where results show that the data spread around the diagonal line and follows the diagonal line's direction, which means the regression model used in this study meets the assumption of normality (Garson, 2012).

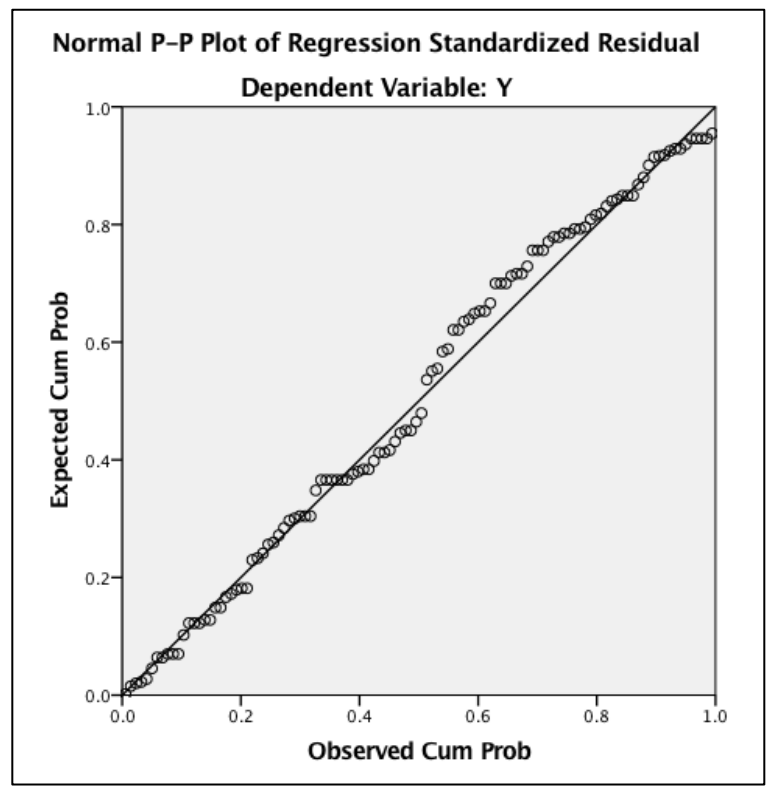

**Figure 3.** P-plot test results show the assumption of normality

In the normality test shown in table 6 with the Kolmogorov-Smirnov Test One-Sample calculation method if the significance value (Sig.) is greater than $p > 0.05$ then the research data is normally distributed according to the results of the analysis that shows ($\beta = 0.136$, $p > 0.05$).

**Table 6.** [One-Sample Kolmogorov-Smirnov Test Results]

| One-Sample Kolmogorov-Smirnov Test | | Unstandardized Residual |
|---|---|---|
| N | | 112 |
| Normal Parameters[a,b] | Mean | .0000000 |
| | Std. | 458.993.107 |
| Most Extreme Differences | Absolute | .076 |
| | Positive | .044 |
| | Negative | -.076 |
| Test Statistic | | .076 |
| Asymp. Sig. (2-tailed) | | .136[c] |
| a. Test distribution is Normal. | | |
| b. Calculated from data. | | |
| c. Lilliefors Significance Correction. | | |

The results of the analysis showed that ($\beta$ = 0.146, p > 0.05), and the conclusion was that there was no heteroscedasticity in the regression model produced in this study. Based on Figure 4, the results of heteroscedasticity testing with scatter plot graphs show that the distribution of points does not form a particular pattern and these points spread randomly above or below point 0 on the y axis, so it can be concluded that in this regression model does not occur heteroscedasticity (Garson, 2012).

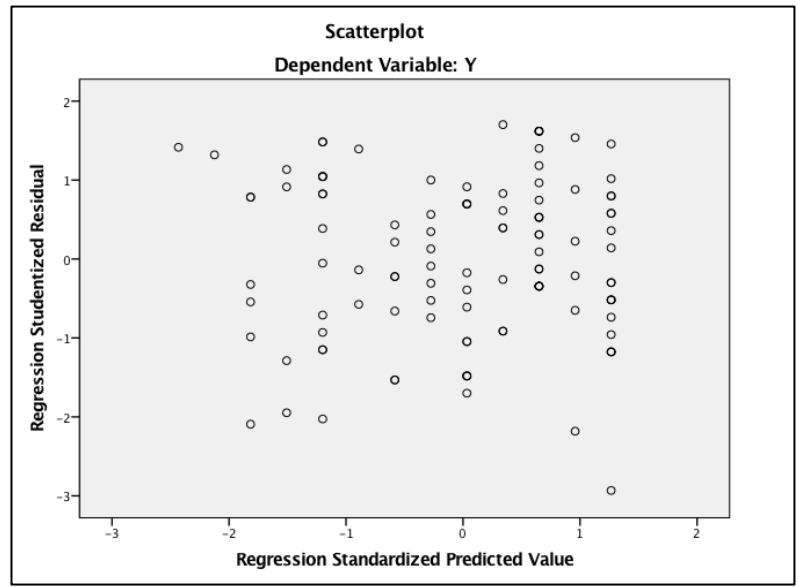

**Figure 4.** [Results of heteroscedasticity on the scatter plot]

The results of the classical assumption test analysis states that all tests were to be significantly fulfilled, including; there is a negative linear relationship between the variable attention (x) with MSB (y), the data are normally distributed, and the absence of heteroscedasticity occurs. A detailed explanation is shown in Table 7.

**Table 7.** [One-Sample Kolmogorov-Smirnov Test Results]

| No | Classic Assumption Test | Results |
|----|-------------------------|---------|
| 1 | Linearity test | Significant negative correlation |
| 2 | Normality test | Meets the assumption of normality. |
| 3 | Heteroscedasticity | No heteroscedasticity |

*4.5 Hypothesis Test Results (T-test)*

In the hypothesis test produced a significance value of p = 0.05, it can be concluded that H0 is rejected and H1 is accepted (Garson, 2012), which means that there is a significant influence of attention (x) toward misinformation sharing behaviour (y). The resulting correlation shows a negative relationship seen from the regression coefficient of (β = -2.894). The complete results of hypothesis testing (t-test) can be seen in table 8.

**Table 8.** [Hypothesis Test Results]

| Coefficients[a] | | | | | | |
|---|---|---|---|---|---|---|
| Model | | Unstandardised Coefficients | | Standardised Coefficients | t | Sig. |
| | | B | Std. Error | Beta | | |
| 1 | (Constant) | 19.533 | 1.053 | | 18.551 | .000 |
| | X | -.390 | .135 | -.266 | -2.894 | .005 |
| a. Dependent Variable: Y | | | | | | |

*4.6 Regression analysis results*

Based on the data analysis that has been generated, the regression equation results are obtained as: (**y=19,533-0,390+e).**

From the regression equation produced in this study, it can be concluded that:

1. The constant value is 19,533, meaning that if there is no change in the attention variable (Value x = 0), the value of misinformation sharing behaviour on social media (y) is 19,533 units.

2. The attention variable's regression coefficient value is -0.390, meaning that if the attention variable increases by 1% and the constant (a) are 0 (zero), then the misinformation sharing behaviour on social media decreases -0.390. Because the regression coefficient value is negative, these results indicate that the variable attention (x) that has a negative correlation toward misinformation sharing behaviour (y), so the higher attention (x), the decreasing misinformation sharing behaviour (y).

The Determination Coefficient (R) analysis in Table 9 shows that the R Square = 0.074. This value implies that the attention (x) on the misinformation sharing behaviour (y) = 7%, while 97% MSB is influenced by other variables not examined in this study.

**Table 9.** [Determination Coefficient Test (R) Results]

| Model Summary[b] | | | | |
| --- | --- | --- | --- | --- |
| Model | R | R Square | Adjusted R Square | Std. Error of the Estimate |
| 1 | .273[a] | .074 | .066 | 4.601 |
| a. Predictors: (Constant), X | | | | |
| b. Dependent Variable: Y | | | | |

# 5. Discussion and limitations

Our analysis revealed that attention is a significant predictor of online misinformation sharing behaviour. Some other contribution variables that are not discussed in this study include the presence of a sense of trust, social influence, cognitive dissonance and epistemic belief, which may also determine factors of misinformation sharing behaviour.

This research aims to investigate the correlation of attentive behaviour towards misinformation sharing behaviour on social media. Our main research problem is based on the understanding that the user is the centre of all efforts, especially concerning psychological factors such as attentive behaviour when understanding the content of the information received. This key attitude is directly related to the dissemination of misinformation. We emphasise that promoting attention behaviour when users are about to share misinformation is an important step in dealing with the spread of fake news and misinformation. The regression analysis results have been found to explain user attention's relationship in reducing (negative correlation) on misinformation sharing behaviour.

Several constraints require to be continued in interpreting the results, namely that this study is based only on an individual survey. Experimental research or data generated by servers from corporate social media platforms may support for further analysis. The findings' generalisability is limited to the research context, which is only limited by attention behaviour in sharing information on social media. This online survey data is only obtained from active users in Malaysia and Indonesia. More data from the cultural or demographic background of some other places is required.

We also conclude that future investigations can study more complex and comprehensive psychological approaches, such as the role of trust factors, epistemic belief, self-actualisation, and social influences to measure further how to deal with misinformation sharing behaviour in online activities. More specifically, future research can measure and collaborate on actual interventions such as designs or tools that serve as stimulants for misinformation-sharing behaviour in applications running on social media platforms.

From the outcome of this research, future research directions include investigating how to improve attention that can influence a person's behaviour in sharing information and a person's approach to interacting with the system and when s/he will share information. The suggested method from the results of this research is to build a system with an awareness of psychological viewpoints such as how to make someone have full attention on the information s/he receives, as s/he receives it.

Having full attention can significantly defeat misinformation sharing behaviour by raising the cognitive abilities maintained by a person. In the end, this can significantly reduce the problems of spreading misinformation that can create a dishonest and untrustworthy belief environment. This approach and its positive influence on the field of Human-Computer Interaction (HCI) needs to be achieved in developing a system, and this is following the results of the research suggested by research (Vosoughi et al., 2018; Bakshy et al., 2009; Cook et al., 2017) which encourage other approaches to the investigation into human judgment factors.

These results indicate that procedures for handling the dissemination of misinformation must also emphasise psychological, behavioural interventions, such as the design of labelling and incentives to limit the expansion of misinformation, rather than concentrating only on restrictions using robotic-based applications. The conclusions recommend that collective disciplines are required in the design of social media applications and information literacy learning to help reduce misinformation sharing and assist users to flag and constructively counter misinformation (Chen et al., 2015).

# 6. Conclusion and future works

Based on this study's findings, we suggest more comprehensive research on the users' psychological role in misinformation. This study's findings can inform a basic theory and empirical research to understand the effects of misinformation on individuals in various domains such as health, security issues such as terrorism, politics, and communication management in crises such as natural disasters. The results from a set of regression analyses confirm that attention factors have a significant negative correlation effect for users to share misinformation on social media, where a higher level of attention (x) significantly reduce misinformation sharing behaviour (y).

This study's findings can also have indirect implications for application developers in designing a proper online social media system based on users' psychological aspects. From the literature that we have examined, it is concluded that the crucial common aspect is a need for a proper understanding of the aspects of human judgment factors. It is expected for us to counter the rate of dissemination of information that is expressly destructive in social media. By comprehensively understanding the phenomenon, research needs a solution collaborating with human factors and what technology to build.

# ACKNOWLEDGEMENT

Thanks also to all the respondents in the study for their involvement. The work was supported by UKM research grant: GPK-4IR-2020-019.